\documentclass[12pt,preprint]{aastex}
\usepackage[usenames,dvips]{color}

\usepackage[usenames,dvips]{color}
\slugcomment{}
\shorttitle{The Central Region of NGC 4945: A Pair of Spirals}
\shortauthors{Lien-Hsuan Lin et al.}

\begin{document}

\title{The Central Region of the Nearby Seyfert 2 Galaxy NGC 4945: A Pair of Spirals}
\author{Lien-Hsuan Lin\altaffilmark{1}, Ronald E. Taam\altaffilmark{1,2}, David C. C. 
Yen\altaffilmark{3}, S. Muller\altaffilmark{4}, and J. Lim\altaffilmark{1,5}}
\altaffiltext{1}{Institute of Astronomy and Astrophysics, Academia Sinica,
P.O. Box 23-141, Taipei 10617, Taiwan, R.O.C.}
\altaffiltext{2}{Department of Physics and Astronomy, Northwestern University, 2131 Tech Drive, 
Evanston, IL 60208}
\altaffiltext{3}{Department of Mathematics, Fu Jen Catholic University, Taipei 24205, Taiwan, R.O.C.}
\altaffiltext{4}{Onsala Space Observatory, SE-43992 Onsala, Sweden}
\altaffiltext{5}{Department of Physics, University of Hong Kong, Pokfulam Road, Hong Kong}
\begin{abstract}
NGC 4945 is a Seyfert 2 galaxy at a distance of 3.82 Mpc.  Its relative proximity has permitted 
a detailed SMA study of the circumnuclear molecular gas in a galaxy exhibiting an AGN.  Based on an analysis  
of the high-resolution velocity field of the central region (20$\arcsec \times$20$\arcsec$, 1$\arcsec$ = 19 pc), we demonstrate that the S-shaped structure of the isovelocity contours is well reproduced by the 
numerical results of a two dimensional hydrodynamical simulation.  In particular, the velocity structure is 
represented by the bending produced by a shock along the spiral density waves, which are excited at the 
outer-inner Lindblad resonance by a fast rotating bar. 
The simulated density map 
reveals a pair of tightly wound spirals in the center which pass through most of the ring-like (claimed to be a circumnuclear starburst ring by other authors) high intensity region in the observations as well as intersect several Pa$\alpha$ emission line knots located outside the ring-like region. 
The calculated mass inflow rate at a scale of 50 pc is about three times the inferred mass accretion rate of the AGN of NGC 4945. We 
find that self-gravity of the gas is important and should be included in our model for NGC 4945. The model 
is compared with the gas orbit model discussed in Lim et al. (2009), and it is shown that the hydrodynamic 
model provides a better match to the observed position-velocity diagram and, hence, provides a more reliable 
prediction of the outer inner Lindblad resonance position.
\end{abstract}

\section{Introduction}
The central kiloparsec region of disk galaxies has received much attention in recent years as a result of 
high-resolution observations obtained by means of mm or sub-mm arrays, the $Hubble$ $Space$ $Telescope$ $(HST)$ 
Near Infrared Camera and Multi-Object Spectrometer (NICMOS), and broad-band optical or near-infrared (NIR) 
images with adaptive optics (AO).  Such capabilities have enabled one to study nearby galaxies on scales as 
small as 10 pc.  Consequently, the nuclear structures of an increasing number of disk galaxies have been 
unveiled. 

In this paper, we undertake a theoretical study of the nearby Seyfert 2 galaxy NGC 4945, which has been classified 
as SB(s)cd or SAB(s)cd (de Vaucouleurs 1964). As it lies at a distance of 3.82 $\pm$ 0.31 Mpc (Karachentsev et al. 
2007), a resolution of 1 arcsec corresponds to a linear scale of 19 pc, thereby making NGC4945 an excellent target for 
high-resolution studies of molecular gas at the center of an active galaxy. 
Among the well-studied AGNs, for example, NGC 1068 (14.4 Mpc), NGC 4258 ($\sim$ 8 Mpc), and M 51 ($\sim$ 8 Mpc), NGC 4945 is closest to us.
  The Seyfert 2 nucleus has been revealed 
by its strong and variable hard X-ray emission (Schurch et al. 2002).  With the NICMOS camera on board $HST$, Marconi et al. 
(2000) identified the structure detected in Pa$\alpha$ emission with a starburst ring of 100 pc in diameter.  Interferometric observations 
of the central region of NGC 4945 in the $^{12}$CO, $^{13}$CO, and C$^{18}$O at their J=2-1 transitions at 
an angular resolution of 5$\arcsec \times$ 3$\arcsec$ were obtained by Chou et al. (2007) with the Submillimeter 
Array (SMA).  They found that all three molecular lines trace an inclined rotating disk with the major axis aligned 
with that of the starburst ring seen in Pa$\alpha$ as well as with that of the large-scale galactic disk. The 
rotating disk exhibits solid-body rotation within a radius of $\sim$ 5$\arcsec$ ($\sim$ 95 pc), 
beyond which the rotation curve flattens. 
The velocity field is marked by a clear S-shaped asymmetry, similar to that revealed by previous HNC observations  (Cunningham \& Whiteoak 2005).  
More recently, 
Lim et al. (2009) reported interferometric observations of the CO (J=2-1) transitions 
at an angular resolution of 2.7$\arcsec$ $\times$ 1.3$\arcsec$ with the SMA. At nearly 
twice the angular resolution of Chou et al. (2007), the S-shaped structure is clearly seen within a radius of 
10$\arcsec$ ($\sim$ 190 pc).  This may be the first time that such a structure is resolved on a 100 pc scale 
near the center of a Seyfert galaxy.

Historically Kalnajs (1978) predicted the existence of an S-shaped structure in the velocity map of the gas in 
a barred galaxy.  For this type of galaxy, the gas is compressed by the bar producing streaming motions on 
either side of a spiral arm.  For a trailing arm, the radial motion of the gas which flows toward the center 
inside corotation and away from center outside corotation, when combined with these streaming motions, leads 
to the bending of the isovelocity curves on the spiral arm and the S-shaped structure in the velocity map.
Such a feature was first observationally seen by Bosma (1978) based on HI  
syntheses data on several 
spiral galaxies and optically by Peterson et al. (1978) in NGC 5383.  Subsequently, it has been confirmed for 
a number of barred galaxies: NGC 1300 (Peterson \& Huntley 1980), NGC 4490 (Duval 1981), NGC 1313 (Marcelin 
\& Athanassoula 1982), NGC 253 and NGC 5236 (de Vaucouleurs et al. 1983), NGC 6221 (Pence \& Blackman 1984), 
NGC 289 and NGC 7496 (Pence \& Blackman 1984), NGC 7741, NGC 3359, and NGC 7479 (Duval \& Monnet 1985), as 
well as NGC 1365 (Lindblad et al. 1996).

In a later theoretical study, Visser (1980) suggested that the S-shaped structure can also be produced in a 
spiral galaxy because a spiral perturbing potential can also produce streaming motions on the gaseous spiral 
arms in a manner similar to that for a bar potential.  Gas dynamical models for the spiral galaxy M81 were constructed 
 demonstrating that the spiral waves can be driven by a spiral potential and the theoretical
S-shaped structures so obtained can match quite well with those present in the  hydrogen-line 
observations.
 
More recently, Yuan \& Kuo (1998) applied their nonlinear asymptotic theory of bar-driven spiral density waves 
(Yuan \& Kuo 1997) to the spiral structure in the central gaseous disks of NGC 1068 and M100.  In their theory, 
two types of S-shaped structures were found.  In one case, the isovelocity contours bend inward along the spiral 
arms and are caused by the radially outward streaming motions across the spiral shocks. This is found for the 
gas flow outside the corotation and provides a fit to the observations of NGC 1068.  On the other hand, the 
second type of S-shaped structure corresponds to outward bending isovelocity contours along the spiral arms 
caused by radially inward streaming motions across the spiral arms. This results from the gas flow inside the 
corotation and can be applied to M100.

The existence of nuclear spirals in barred galaxies has also been shown in some hydrodynamical simulations.
For example, Athanassoula (1992) pointed out that the curved inner parts of offset shocks along the bars curl around the nucleus and their form is very reminiscent of the often-observed nuclear spirals.
Good examples are NGC 4314 and NGC 1097.
Maciejewski (2004b) mentioned that any asymmetry in the nuclear potential may generate nuclear spirals inside its inner Lindblad resonance.

Based upon these theoretical results, the clear S-shaped structure of the isovelocity contours in the 
CO (J=2-1) velocity map provides strong evidence for the presence of spiral density waves in the nuclear 
region of NGC 4945.  In order to obtain a better understanding of the structure and kinematics of the nuclear 
region, we report on results from two-dimensional hydrodynamical simulations of spiral density waves which 
are compared to the CO (J=2-1) velocity map as well as the Pa$\alpha$ image of Marconi et al. (2000).  

The numerical method and procedure underlying our model are discussed in \S 2.
In \S 3, we present our simulation results and compare them with observational features of NGC 4945 as well as results obtained by Lim et al. (2009) using their gas orbit model.
Finally, we summarize and conclude our work in the last section.

\section{Numerical Method}

The hydrodynamic simulations were performed with a high-order Godunov code known as Antares, in which the 
hydrodynamic fluxes on zone interfaces are calculated with the exact Riemann solution (Yuan \& Yen 2005). 
Since the central region of this galaxy is of primary interest, Cartesian coordinates are adopted in the 
present computations in order to avoid the need to impose an artificial inner boundary condition to handle 
the coordinate singularity at the origin. At the outer boundary, radiation boundary conditions are imposed.  
That is, the wave characteristic decomposition is performed at each boundary cell, allowing waves to propagate 
outward, but suppressing incoming waves. In this treatment, reflection is not permitted at the boundary. 

We consider the evolution of the gaseous disk in the isothermal approximation.  The equations governing 
the flow corresponding to mass conservation and motion are:  
\begin{eqnarray}
\frac{\partial \sigma}{\partial t}+\nabla \cdot (\sigma \textbf{\textit{v}}) =0, \\
\frac{\partial \textbf{\textit{v}}}{\partial t}+\textbf{\textit{v}}\cdot \nabla \textbf{\textit{v}}
=-\frac{\nabla P}{\sigma}-\nabla V,
\end{eqnarray}
where $\sigma$ denotes the surface density of gas in the disk and $\textbf{\textit{v}}$ denotes the velocity 
vector. Here, $P$ is the gas pressure integrated in the $z$-direction. For an isothermal gas,
\begin{equation}
P = a^2\sigma,
\end{equation}
where $a$ is the sound speed. The total gravitational potential, $V$, is composed of three components:
\begin{equation}
V = V_0+V_1+V_g.
\end{equation}
The first term, $V_0$, is a central potential supporting a differentially rotating disk, i.e.,
\begin{equation}
\frac{dV_0}{dr}=r\Omega^2(r),
\end{equation}
where the angular speed $\Omega(r)$ is determined from the observed rotation curve.  The second term, $V_1$, 
is the rotating non-axisymmetric potential of interest. For the purpose in this paper, we adopt a potential from 
a stellar bar rotating as a rigid body as 
\begin{equation}
V_1(R,\phi,t)=\Phi(R)\cos[2(\phi-\Omega_pt)],
\end{equation} 
where $\Omega_p$ is the angular speed of the bar and only the first non-vanishing harmonic term is included. 
Note that the axisymmetric part of the bar potential is already contained in $V_0$.  A simplified functional 
form for the amplitude is adopted as:
\begin{equation}
\Phi(R)=-\Phi_0\frac{R^2}{(A^2_1+R^2)^2},
\end{equation}
where $A_1 \equiv a_1/r_s$ , $R \equiv r/r_s$, $r_s$ = 1.0 kpc, and $\Phi_0$ is a parameter related to the strength of the bar potential. 
The functional dependence on the radial coordinate $R$ is chosen to preserve the asymptotic behavior of 
the potential.  It is proportional to $R^2$ as $R\rightarrow 0$ and approaches $R^{-2}$ for large $R$.  Thus, 
the force of the bar at $R=0$ is zero, while it behaves as $R^{-3}$ when $R$ is large. The parameter $a_1$ 
denotes the radial distance at which the bar potential is a minimum.  Finally, $V_g$, the last term of $V$, 
represents the self-gravitational potential of the gaseous disk.

In order to include the 
self-gravity of the disk in the calculation, the hydrodynamic code is coupled to a Poisson solver of kernel representation.
  The Poisson equation for the disk is intrinsically 
three dimensional; however, the solution of the complete potential function problem is avoided.  
Specifically, the force is calculated by integration in the plane of the disk where the force can be written 
as a double summation of the product of surface density and a kernel, which is in the form of a convolution 
(Yen et al. 2011, in preparation).  The computational demand associated with the convolution of two vectors of length $N$ 
is reduced from $O$($N^2$) to $O$($N$ln$N$) with the FFT. As a result, the entire force computation has a demand 
of $O$($N^2$(ln$N$)$^2$).  Since the FFT is only used to accelerate the computation, our Poisson solver 
does not require the imposition of a periodic boundary condition.

\subsection{Computational Setup}
The initial surface density of the gas is taken to be uniform with a value of 153 M$_\sun$ pc$^{-2}$, which is estimated from the hydrogen gas mass by Chou et al. (2007) 
based upon the $^{12}$CO (J=2-1) line emission. Specifically, the mass in molecular hydrogen gas is 
estimated as (1.63 $\pm$ 0.03) $\times$ 10$^{8}$ M$_\sun$ for the disk of 16.4$\arcsec \times$ 10.8$\arcsec$ 
($\pm$0.1$\arcsec$) ($\sim$ 310 $\times$ 205 pc) (at FWHM).

\begin{figure}
\figurenum{1}
\epsscale{1.0}
\plotone{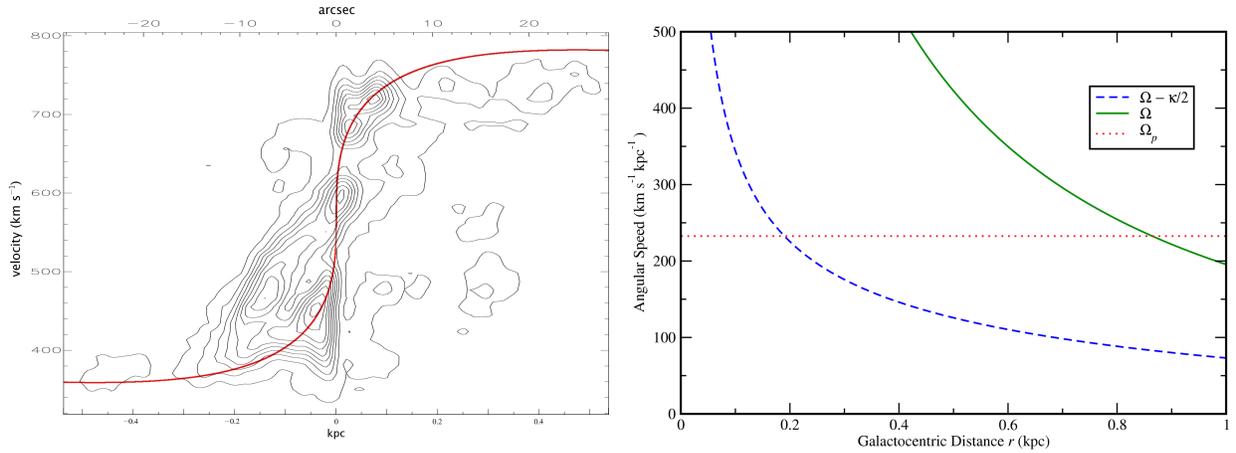}
\caption{Left panel: adopted representation of the rotation curve of NGC 4945 ($red \ solid \ line$) superposed on the observed $^{12}$CO (J=2-1) P-V diagram in the left panel of Fig. 9 in Lim et al. (2009)
for the axisymmetric force in our model. 
Right panel: angular speed curves as a function of radius derived from the rotation curve, where $\Omega$ and $\kappa$ are the circular angular speed and the radial epicyclic frequency respectively.
The horizontal lines represent the pattern speed of the bar, $\Omega_{p}$, which is 233 km s$^{-1}$ kpc$^{-1}$ for the gaseous disk with self-gravity.
The intersection of $\Omega_{p}$ with the $\Omega - \kappa$/2 curve determines the location of the inner Lindbald resonance (ILR) to be 0.19 kpc.
}
\label{rcagd}
\end{figure}

For the axisymmetric background force, we use as initial input the representation of the rotation curve of NGC 4945 illustrated by the solid line
in the left panel of Figure~\ref{rcagd}.  
It is in the form of an Elmegreen rotation curve (Elmegreen \& Elmegreen 1990) 
written as
\begin{equation}
v(r)=v_0(\frac{r}{r^B+r^{1-A}}).
\end{equation}
This form is a general representation of rotation curves for spiral galaxies and is characterized by a nearly 
rigid-body velocity profile at small $r$ and a nearly flat profile at large $r$.  The optimal values 
of $v_0, B,$ and $A$ for NGC 4945 are found to be 391 km s$^{-1}$, 0.661, and $-0.772$ respectively that provide the best match with the observed P-V diagram for NGC 4945.  
The corresponding angular speed curves are shown in the right panel of Figure~\ref{rcagd}  where the horizontal line represents $\Omega_p$, the 
pattern speed of the bar (see \S2.2 for its determination).  The intersection of $\Omega_p$ with the $\Omega - \kappa/2$ curve determines the 
location of the outer inner Lindblad resonance (OILR).  Here, $\kappa$ is the epicyclic frequency.  Since 
$\Omega - \kappa/2$ increases monotonically when $r$ decreases, an inner inner Lindblad resonance (IILR) does not 
exist.

To start the simulation, the gas is placed on circular orbits with rotational velocities given by the 
initial rotation curve. In order to reduce numerical noise, the amplitude of the bar potential, $\Phi_0$, is increased 
gradually, growing linearly from zero to its full value during one rotation period of the bar.  This procedure 
and the amount of time required for the start up was found to be necessary to avoid the formation of transients 
resulting from the growth of the perturbing potential.

To complete the specification of the input parameters, we adopt a sound speed, $a$, for the gaseous disk 
of 10 km s$^{-1}$ (Lindblad et al. 1996; Lindblad \& Kristen 1996) corresponding to a temperature $\sim 10^4$ K.
The gas response is not very sensitive to this parameter within about 5 $\sim$ 15 km s$^{-1}$ in our tests.
Patsis \& Athanassoula (2000) also showed that the sound speed makes little difference to the gas density from 10 to 15 km s$^{-1}$.
The computational domain corresponds to a physical region of 0.8 kpc $\times$ 0.8 kpc. 
Since the calculations presented in this paper are performed with a 512 $\times$ 512 uniform grid, the cell size is 1.56 $\times$ 1.56 pc$^2$ (or 0.039$\arcsec \times$ 0.039$\arcsec$).

\subsection{Free Parameters}
The free parameters important for the gas response to the rotating bar are the angular pattern speed, the location 
of the potential minimum, and the strength of the bar potential.  These parameters are varied to seek the best-fit 
models.

The pattern speed $\Omega_p\,$ controls the positions of the resonances and, thus, the morphology of the spiral arms.
We have chosen to take as an initial estimate of the pattern speed a value that places the outer inner Lindblad 
resonance (OILR) outside the ring-like feature with an outer radius of $\sim$ 4$\arcsec$ ($\sim$ 80 pc) (Lim et 
al. 2009) seen in the Pa$\alpha$ image.  
Calculations based on the linear theory of the gas response to an asymmetric potential (Maciejewski 2004a) showed that the long spiral density waves generated at the OILR propagate outwards and get reflected as short spiral density waves at the boundary of the penetration zone.
These short waves then freely propagate inwards through the resonance 
and their radial wavelength decreases during the process.  As a result, the spiral arms become more tightly 
wound and take the appearance of a ring-like morphology around the center.  
Furthermore, it has been shown in hydrodynamical simulations that nuclear spirals or nuclear rings are formed inside the OILR (Athanassoula 1992; Piner et al. 1995; Maciejewski 2004b).
An analysis of the central region in 
M100 (NGC4321) by Knapen et al. (1995) also shows that the nuclear ring should be located inside the OILR.
$\Omega_p$ takes a value between 213 $\leq \Omega_p \leq$ 249 km s$^{-1}$ kpc$^{-1}$, corresponding to a radial 
interval for the radius of the OILR of 0.17 kpc $\leq$ R$_{OILR}$ $\leq$ 0.22 kpc.

The location of the minimum of the bar potential, a$_1$ in eq. 7, is taken to be in the range 0.08 $\leq$ a$_1\leq$ 
0.12 kpc,  so that it also lies outside the ring-like feature.  As described in Lin et al. (2008), the nuclear ring 
forms inside the location of the potential minimum.  To parameterize the strength of the bar potential, we denote 
by $f$ the ratio of the effective radial force exerted on the disk by the rotating bar (Yuan \& Kuo 1997) to the 
force for the circular motion from the rotation curve at the location of OILR.  This choice is motivated by the 
fact that the waves are excited at the resonances.  It is found that $f$ should lie between 1.5\% and 2.05\%. 

\section{Results}
In order to compare our theoretical results with the observations, the plane of the theoretical disk must be 
oriented with respect to the line of sight. From the analysis of the mean velocity map of $^{12}$CO (J=2-1) of 
the central region of NGC 4945 (the upper-left panel of Figure~\ref{vf}), the P.A. of the major axis is 48$^\circ$ and the P.A. of the 
line of nodes is 20$^\circ$. The inclination for the nuclear disk is 62$^\circ$ (Chou et al. 2007), somewhat 
smaller than the inclination of the large-scale galactic disk of $\sim$ 78$^\circ$.

\subsection{Comparisons between numerical results and observations}
The main purpose of this study is to obtain an understanding of the nuclear region of NGC 4945 by simulating as much as possible the observed features using a simple yet physically plausible model.
We use the same method employed by Lindblad et al. (1996) and Lindblad \& Lindblad (1996) to compare the simulations with observations.
We judge our models by visual inspection of the model density and velocity maps overlaid on the observed ones, instead of adopting a quantitative method, e.g. $\chi^{2}$, which can be misleading if the strong density features or steep velocity gradients present in the models do not exactly overlap their observed counterparts. Such comparisons would be sufficient in the context of our simple model.
Furthermore, since our model is symmetric, it is not our goal in this study to simulate complex and asymmetric features in observations.

To obtain the best-fit model, we have carried out 30 simulations in total. 
The free parameters were varied during the search for the best-fit model in the following manner.
According to the dispersion relation (equation 16 in Maciejewski 2004a), the morphology of the spiral arms is mainly determined by the epicyclic frequency, i.e. the rotation curve, and the pattern speed of the bar potential, i.e. the positions of the resonances.
Since we use the same rotation curve in all our simulations, the most important parameter for obtaining the best-fit model is the pattern speed of the bar potential.
Based on the Pa$\alpha$ image, we varied the position of the OILR between 0.17 $\leq$ R$_{OILR}$ $\leq$ 0.22 kpc.
The other two free parameters were adjusted accordingly.

The best-fit values of the parameters to the observations 
are $\Omega_p$ = 233 km s$^{-1}$ kpc$^{-1}$, a$_1$ = 0.09 kpc, and $f$ = 1.8\% (see Table 1).  The inferred value 
of $\Omega_p$ implies that the position of the outer inner Lindblad resonances (OILR) is located at R$_{OILR}$ = 
0.19 kpc.  The corotation resonance (CR) and the outer Lindblad resonance (OLR) both lie outside the computational 
domain (-0.4 kpc $\sim$ 0.4 kpc along both the x and y axes).  According to eqs. 6 and 7, a$_1$ denotes the length 
scale of the bar and, therefore, the perturbing potential determined by these best-fit values is that of a 
small fast-rotating weak bar.

\subsubsection{velocity fields}
\begin{figure}
\figurenum{2}
\epsscale{0.7}
\plotone{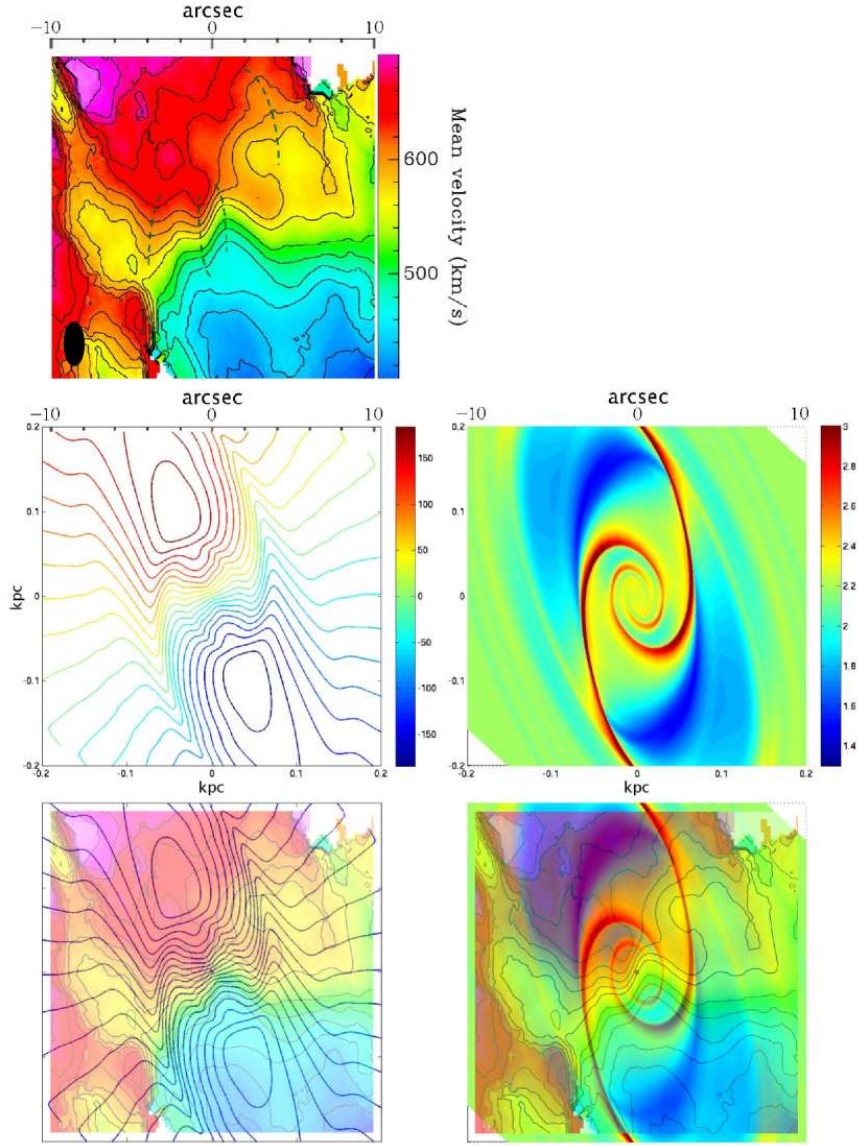}
\caption{
Upper-left: intensity-weighted mean velocity map of $^{12}$CO(2-1) of the central region (20$\arcsec \times$20$\arcsec$) of the galaxy observed with the SMA. 
The map shows a velocity gradient from redshift to blueshift, as indicated by the color bar.
The synthesized beam is shown in the lower left corner, and has a size at FWHM of 2.72$\arcsec \times$1.27$\arcsec$ with a position angle of 1$^{\circ}$.
In the central $\pm$8$\arcsec \times \pm$8$\arcsec$ region, the isovelocity contours show several sets of curved lines marked with green dashed lines, 
which imply the locations of spiral density waves.
North is at the top and east is to the left.
Middle panels show the projected simulation results of the gaseous disk with self-gravity. 
Middle-left: isovelocity contours of the simulated velocity field convolved with the synthesized beam in the upper-left panel. 
The unit on the color map is km/s.
Middle-right: the simulated density distribution. 
The color map denotes the surface density distribution in logarithmic scale in units of M$_\sun$ pc$^{-2}$.
Lower-left: the superposition of the middle-left panel onto the upper-left panel.
Lower-right: the superposition of the middle-right panel onto the upper-left panel.
The length scales of these five plots are the same.
}
\label{vf}
\end{figure}

The intensity-weighted mean velocity map of $^{12}$CO (J=2-1) obtained with the SMA by Lim et al. (2009) is 
displayed in the upper-left panel of Figure~\ref{vf}.  In the central $\pm$8$\arcsec$ $\times \pm$8$\arcsec$ region, several sets of curved 
lines (representing a range of isovelocity contours) are marked with green dashed lines, which indicate the locations 
of the purported spiral density waves.
Although we cannot read from this map the intensity of the emission, by looking at the observed data in general we can see that the emission outside the $\pm$8$\arcsec \times \pm$8$\arcsec$ region is much weaker than that inside.
Therefore, the reliable part of the observed velocity field is in the $\pm$8$\arcsec \times \pm$8$\arcsec$ region and we will focus on features in this region.

The projected isovelocity contours and surface density distribution of the
best-fit simulation result are illustrated in Figure~\ref{vf} and compared with
observations.
In the lower-left panel of Figure~\ref{vf}, the sets of curved lines indicated by the leftmost and rightmost green dashed lines in the observed velocity map are reproduced in the simulated velocity field.
At the same time, two spirals pass through these two sets of outward bends in the lower-right panel of Figure~\ref{vf}. 
In this case, the outward bends 
face the inward propagating spiral with their convex side.  This is one of the distinctive features for the density 
waves excited at the OILR (Kalnajs 1978, Yuan $\&$ Kuo 1998).  Across the spirals, matter collides with the arm 
from the concave (or inner) side and is deflected by the shock, resulting in radially inward streaming motions and 
mass inflow, the rate of which we calculate in \S 3.2.
However, the bending features indicated by the two green dashed lines around the galactic center in the observed velocity map are dissimilar to those in the simulation.
It is noticeable that these two green dashed lines span an area roughly the same size as that of the synthesized beam in the observed velocity map.
Observations of even higher resolution might be needed to make it feasible to compare the simulation with observation in this region.

\begin{figure}
\figurenum{3}
\epsscale{1.0}
\plotone{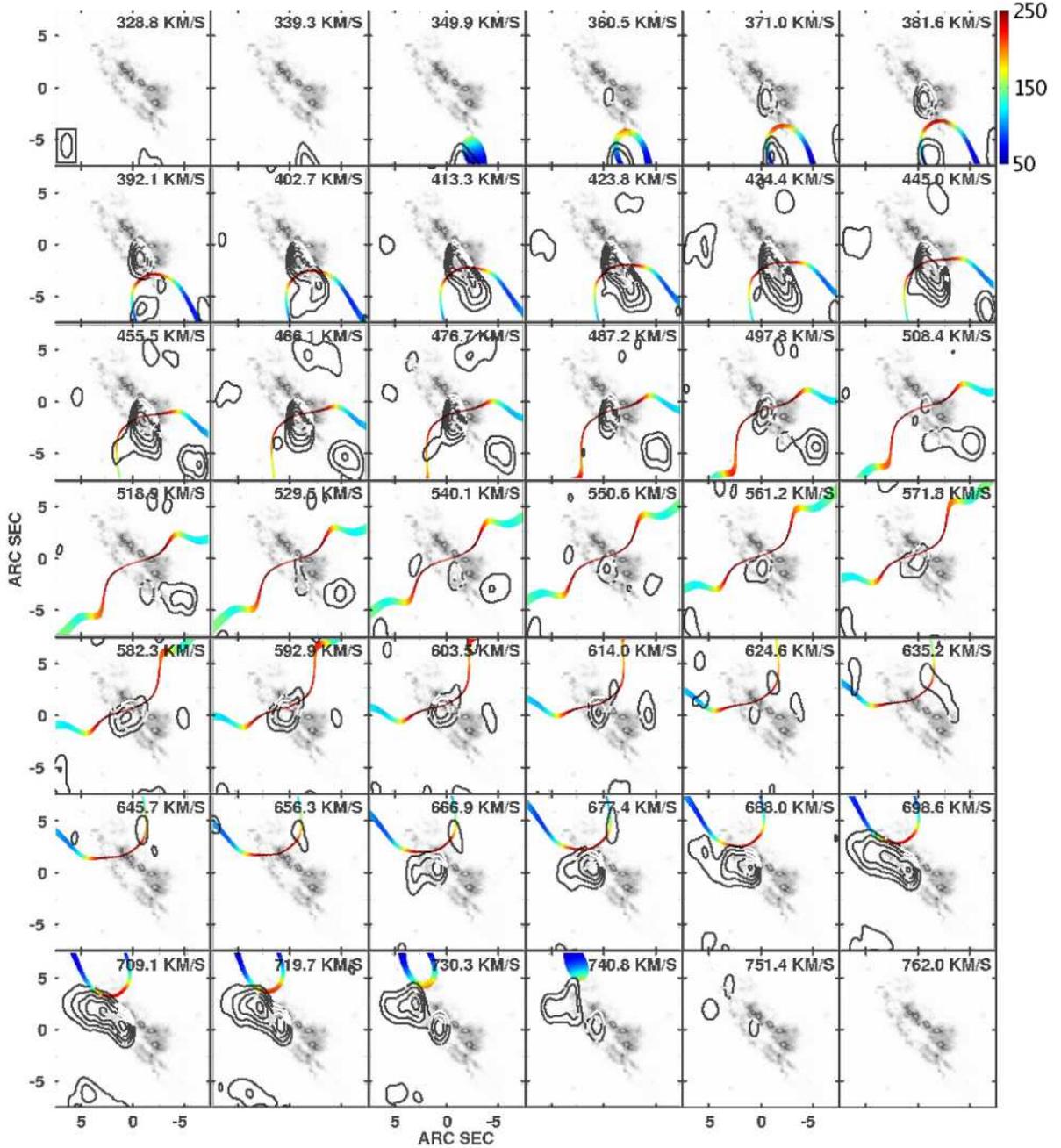}
\caption{
$^{12}$CO(J=2-1) velocity channel maps in contours overlaid on the Pa$\alpha$ image (grey scale) from Marconi et al. (2000) and the simulated channel maps (color scale). Contour levels are plotted at 5, 10, 16, 23, 31, 40, 50, and 61 $\times$ 92 mJy beam$^{-1}$ (rms noise level).
Velocity increases from the upper left channel to the lower right channel. 
The synthesized beam, shown in the lower left corner of the top left panel,  has a size at FWHM of 2.32$\arcsec$ $\times$ 1.07$\arcsec$ at position angle -1.1$^{\circ}$. 
The color map indicates the surface density on a linear scale in units of M$_\sun$ pc$^{-2}$
.}
\label{chnmp}
\end{figure}

The observed and simulated velocity channel maps are shown in Figure~\ref{chnmp}.
The $^{12}$CO (J=2-1) channel maps obtained from the high angular resolution interferometric observation obtained with the SMA by Lim et al. (2009) are drawn in contours, while the color maps are the simulated velocity channel maps with the color bar indicating the surface density on a linear scale.
For the observed channel maps,
the emission appears first in the south-west and gradually moves closer to the center, from the low velocity channels to the high velocity channels, intensifying at first before 
weakening.  After passing through the center, it continues to move to the north-east, regaining strength before 
weakening again.
The same tendency can also be found in the simulated channel maps only the weakening of the central part is not as obvious as in the observed ones.  
The highest density regions in the simulated channel maps and the strongest emissions in the 
observed ones occur at similar positions.  

\begin{figure}
\figurenum{4}
\epsscale{1.0}
\plotone{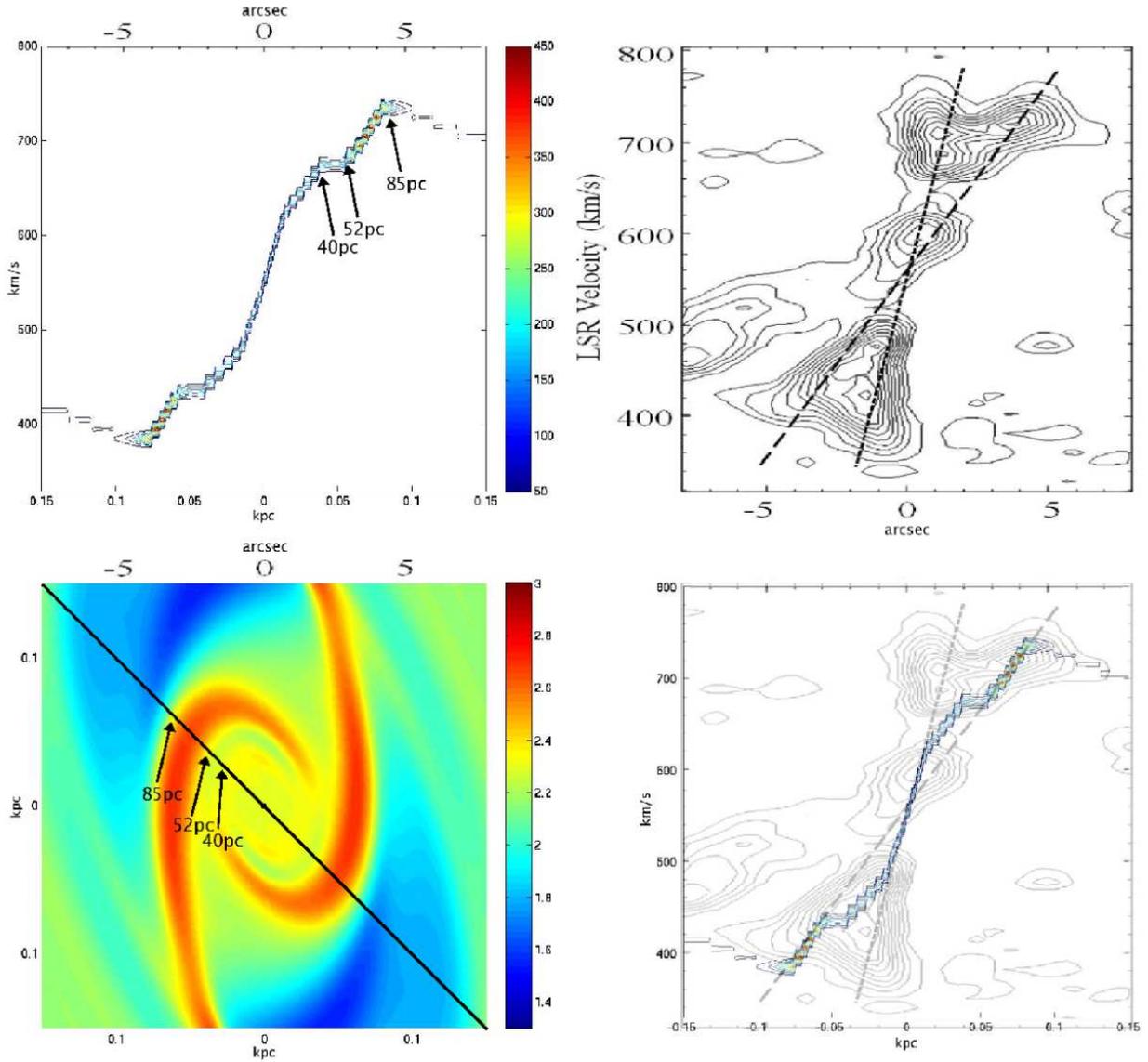}
\caption{
Upper-right: the observed P-V diagram in $^{12}$CO (J=2-1) in the left panel of Fig. 9 in Lim et al. (2009)
in which they use two dashed lines to indicate two distinct kinematic components.
1$\arcsec$ = 19 pc. Upper-left: the simulated P-V diagram along the line at P.A. = 45$^\circ$. The color map denotes the surface density on a linear scale in units of M$_\sun$ pc$^{-2}$. 
Lower-right: superposition of the two upper panels.
Lower-left: the projected simulated density map convolved with the synthesized beam.
The color map indicates the surface density in logarithmic scale in units of M$_\sun$ pc$^{-2}$.
}
\label{pv}
\end{figure}

The observed P-V diagram in $^{12}$CO (J=2-1) (left panel of Figure 9 in Lim et al. (2009)) is 
reproduced in the upper-right panel of Figure~\ref{pv}.  Lim et al. (2009) used two dashed lines with different 
velocity gradients to indicate two distinct kinematic components.  They interpreted the feature with the steeper 
velocity gradient as a compact circumnuclear molecular disk having an outer radius of $\sim1\arcsec$ ($\sim$20 
pc), and considered the possibility that this disk may correspond to the hypothetical circumnuclear molecular 
disk/torus invoked by AGN unification models.  The other feature with the shallower velocity gradient, which is 
coincident with and extends somewhat beyond the Pa$\alpha$ starburst ring, was identified with a circumnuclear 
molecular ring having an outer radius of $\sim4\arcsec$ ($\sim$80 pc).  The upper-left panel of Figure~\ref{pv} is 
the simulated P-V diagram along the line at P.A. = 45$^\circ$ with the densest region located at 
50-90 pc and the second densest region located at 0-40 pc.  
In the lower-right panel of Figure~\ref{pv}, which is a superposition of the simulated and observed P-V diagrams, it can be seen that
 the position of the peak of the densest region in the simulation coincides with the corresponding position in the observation.  
The projected simulated density map (the middle-right panel of Figure~\ref{vf}) convolved with the observed synthesized beam is shown in the lower-left panel of Figure~\ref{pv}.  
After convolution, the original spirals seem to separate into two components: 
a dense ring if one neglects the extended outer arms and a less dense circumnuclear disk.  
The densest region along the line is at 50-90 pc and the second densest region is inside 40 pc which are 
consistent with the simulated P-V diagram.  
However, the central part of the circumnuclear disk is less dense than its outer region since the amplitude of the spiral waves is attenuated during inward propagation then the density of the spirals becomes lower.
Based on these comparisons, it can be seen that the 
 compact and ring-like features at the centers of galaxies found in observations may just be unresolved spiral features.  
Therefore, instead of a circumnuclear molecular torus, tightly wound spiral arms may provide an alternative 
interpretation for such a feature surrounding an AGN on this spatial scale.  The only feature in the observed 
P-V diagram 
not reproduced in our simulation result is the higher velocity peak located at $\pm$ 1$\arcsec$, which appears on 
both the red-shifted and blue-shifted halves of the observed P-V diagram.  
Some of the P-V diagrams in Figure 3 in Athanassoula \& Bureau (1999) have shown an inverted S-shaped feature (not the same S-shaped structure in the isovelocity contours in this paper) which is due to the x$_2$ orbits of a simple bar potential (Bureau \& Ahtanassoula 1999).
For example, in the P-V diagram at $\Psi$ = 22.5$^{\circ}$, the velocity distribution at D = 0.5 kpc may correspond to two peaks at the same position in an observed P-V diagram.
However, the position of the two peaks in the observed P-V diagram in Figure~\ref{pv} of this paper is between 20$\sim$30 pc, which is in the innermost region of NGC 4945 where the orbits are nearly circular.
According to Bureau \& Athanassoula (1999), the trace of a circular orbit in a P-V diagram is an inclined straight line passing through the origin and identical for all viewing angles.
Therefore, the reason underlying the existence of the two peaks at the same position at such a small scale remains unclear.

\subsubsection{Images}
\begin{figure}
\figurenum{5}
\epsscale{1.0}
\plotone{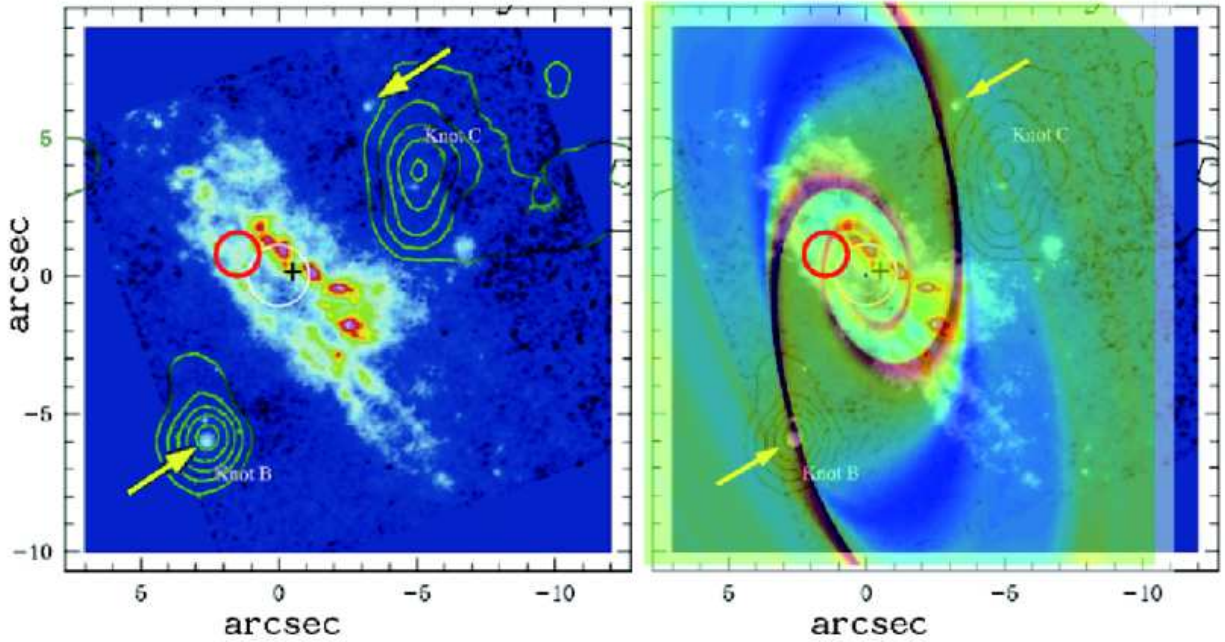}
\caption{
Left panel: continuum-subtracted Pa$\alpha$ image (Marconi et al. 2000). North is up and East is left.
The cross marks the nucleus of the K-band image and the white circle represents the uncertainty on the position of the H$_{2}$O maser given by Greenhill et al. (1997).
The contours (denoted by Knot B and Knot C) are from the H$\alpha$+[NII] image by Moorwood et al. (1996).
The red circle highlights an emission which is not part of the strong ring-like emissions.
Two emission line knots located outside the ring-like region are marked by arrows.
Right panel: the superposition of the middle-right panel of Fig.~\ref{vf} onto the left panel. 
The spirals in our simulation pass through most of the strong emission line knots surrounding the center, 
two emission line knots (marked by arrows) located outside the ring-like region,
as well as an emission (circled in red) which is not located on the ring-like region.
}
\label{image}
\end{figure}

In the left panel of Figure~\ref{image}, the continuum-subtracted Pa$\alpha$ image (Marconi et al. 2000) is illustrated. Besides 
the strong central emission, two emission line knots (marked by arrows) are located outside the ring-like region. 
We note that the emission highlighted by a red circle lies next to the white circle, which represents the 
uncertainty in the position of a $H{_2}O$ maser, that is not located on the ring-like region.

The right panel of Figure~\ref{image} shows the superposition of the Pa$\alpha$ image and the simulated projected surface density distribution (the 
middle-right panel of Figure~\ref{vf}).  
It can be seen that 
there is a pair of tightly wound spirals in the central region in the simulated image, in distinct contrast to the feature previously 
described as a circumnuclear (starburst) ring.
The spirals in our simulation not only pass through most of the strong 
emission line knots surrounding the center, but also intersect with two emission line knots (marked by arrows) 
located outside the ring-like region and about symmetric to the center. 
Especially the emission Knot B is detected both in the Pa$\alpha$ image by Marconi et al. (2000) and the H$\alpha$ + [NII] image by Moorwood et al. (1996).
Nevertheless, some emission knots which are outside the ring-like region, e.g. the southwestern part to the center, in the Pa$\alpha$ image do not lie on the spirals of our model.
In general, such more complex and asymmetric features in the observation are difficult to simulate within the framework of our simple model.
Comparing the right panel of Figure~\ref{image} with the lower panels of 
Figure~\ref{vf}, it can be seen that two emission knots (denoted by Knot B and Knot C) in the H$\alpha$ + 
[NII] image by Moorwood et al. (1996) are (i) coincident with the bendings of the isovelocity curves and (ii) the extended 
emission of Knot B is aligned along the simulated spiral arm.  Since the observation shows that the structure is 
not symmetric about the center, it is natural that the position of Knot C is shifted slightly from that of the 
simulated spiral arm.  Meanwhile, the spiral entering from the north and to the west of the center, which makes a 
turn at the south towards the center, passes through the emission circled in red in Figure~\ref{image}.
Although this emission is not strong, the coincidence of the positions of this emission knot and the turning of the spiral in the simulated map may point to the possible presence of a pair of spiral arms in the central region of NGC 4945.

\begin{figure}
\figurenum{6}
\epsscale{1.0}
\plotone{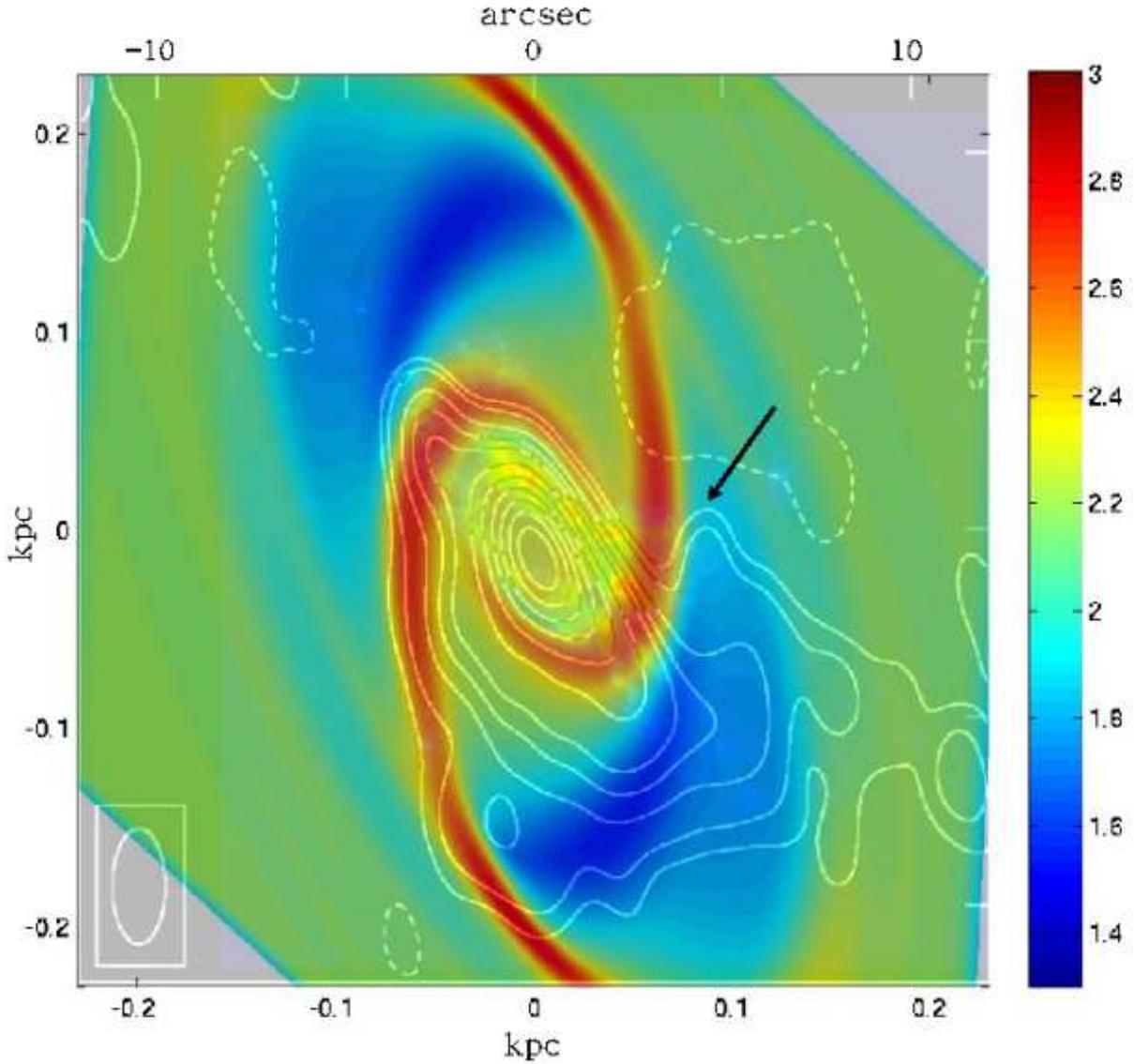}
\caption{
The contours, reproduced from the left panel of Fig. 1 in Lim et al. (2009), represent the 1.3 mm continuum emission.
North is up and East is left.
The extended emission to the west of the center is marked by an arrow.
The color map denotes the simulated surface density distribution which is the middle-right panel of Fig.~\ref{vf} convolved with the synthesized beam in the lower-left corner. 
The color bar is in logarithmic scale in units of M$_\sun$ pc$^{-2}$. 
}
\label{continuum}
\end{figure}

The simulated projected surface density distribution (the middle-right panel of Figure~\ref{vf}) convolved 
with the synthesized beam in the 1.3 mm continuum contour map (the left panel of Figure 1 in Lim et al. 2009) 
and superposed onto the 1.3 mm continuum contour map 
is shown in Figure~\ref{continuum}. It is notable that the extended emission west of the center 
(marked by an arrow) in the observation follows the spiral arm in the simulation in spite of a slight shift in 
position.  Although strictly speaking the 1.3 mm continuum map traces the column density, if the gaseous disk is 
thin, there is little difference between column and surface density.  Moreover, a similar extended emission at 
the same position as that in the 1.3 mm continuum map is also seen in the $^{12}$CO (J=2-1) integrated intensity 
map shown in the upper-left panel of Figure 4 in Lim et al. (2009).
Nevertheless, some extended emission in the southwestern part is an asymmetric feature which is not simulated in our simple model.

\subsection{Mass inflow rate}
The idea of fueling AGNs through OILR-excited spiral waves is well known and is thought to occur in the case of a 
fast rotating bar in a galaxy with a rapidly rising rotation curve. In this case, the OILR is located within a few 
hundred parsecs from the center, and there may not exist an IILR.  The trailing spiral density waves excited at 
OILR will propagate inward and the amplitude of the waves will be attenuated by viscosity.  The negative angular 
momentum carried by the waves will be deposited over the annular region traversed by the waves, leading to the flow 
of disk material, after losing angular momentum, toward the center.

\begin{figure}
\figurenum{7}
\epsscale{1.0}
\plotone{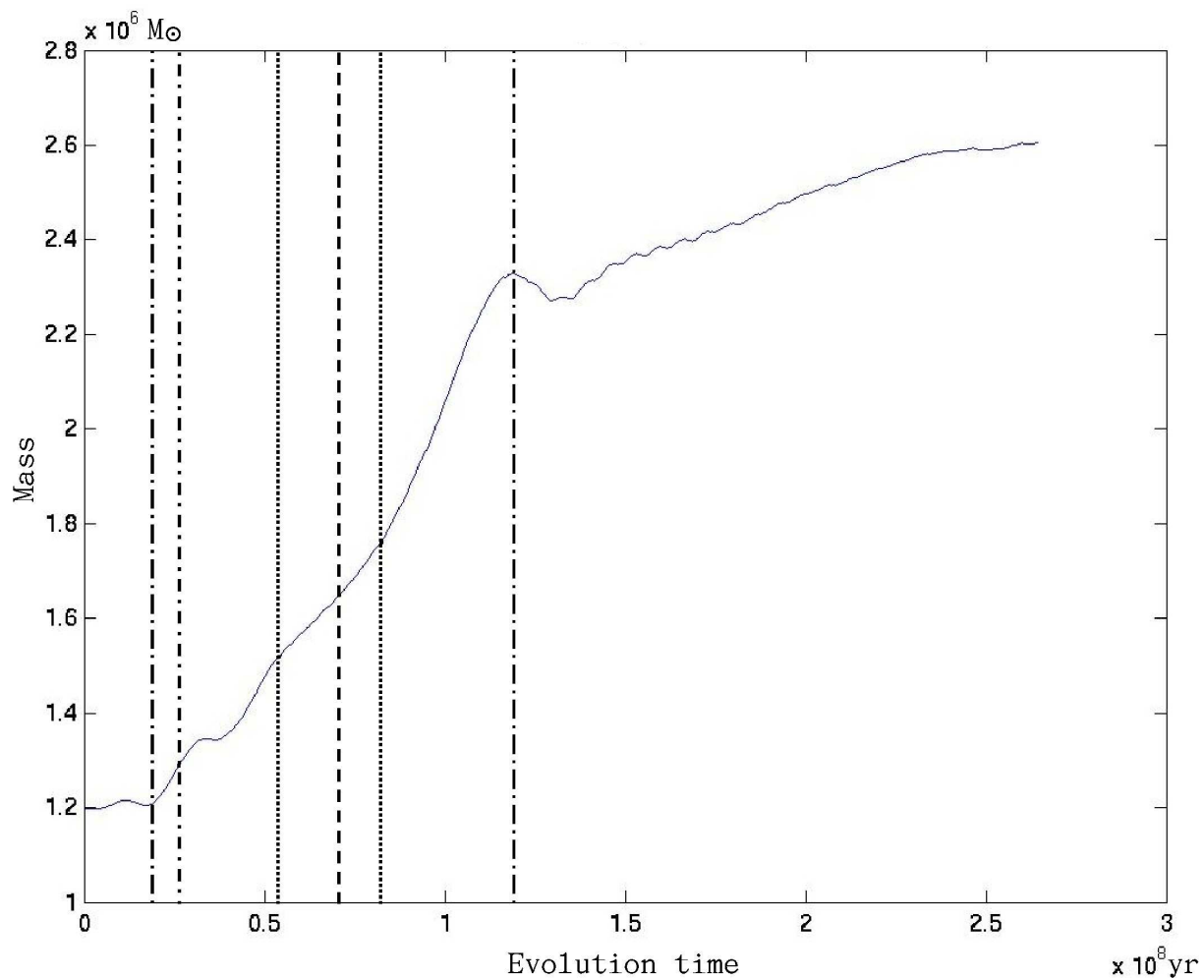}
\caption{
Total mass in M$_\sun$ within a radius of 50 pc as a function of time in years.
The two dotted-long-dashed lines roughly separate the curve into three distinct periods with different mass inflow rates.
The amplitude of the bar potential is increased gradually, growing linearly from zero at the beginning to its full value at one revolution of the bar indicated by the dotted-dashed line. 
The dashed line indicates the time we pick the simulation result to compare with observations. 
We estimate the mass inflow rate by the slope of the total mass curve during the period between two dotted lines.
}
\label{mass}
\end{figure}

We plot the total gaseous mass within a radius of 50 pc as a function of time in Figure~\ref{mass}.  
Two dotted-long-dashed lines roughly separate the curve into three distinct segments, corresponding to three different periods with different mass inflow rates.
The first period is from zero to 20 Myr (indicated by the leftmost dotted-long-dashed line in the figure).
The mass inflow rate is very low during this period since the amplitude of the bar potential is in the process of being increased gradually from zero to nearly its full value and the shocks along the spirals have not yet been formed.
The second period is between two dotted-long-dashed lines.
The mass inflow rate is the highest during this period since strong shocks have formed and mass in the outer bar region is accumulating into the central region.
The third period is after 120 Myr.
The slope of the curve becomes shallow again since the morphology inside the radius of 50 pc has reached the quasi-steady state and less mass is left in the outer bar region for accretion into the central region.
The best-fit 
frame to the observations is obtained at a time of 71 Myr.  
We estimate the mass inflow rate by the slope of the total mass curve during the
period between the two dotted lines, finding a rate of 0.0085 M$_\sun$ yr$^{-1}$ at 50 pc.  Since 50 pc lies inside the ring-like starburst 
region, the gas flowing into the 50 pc region should be responsible for fueling the AGN rather than for feeding a star burst.

According to Guainazzi et al. (2000), the nuclear luminosity of NGC 4945 is 1.77 $\times$ 10$^{43}$ erg s$^{-1}$.
If the central engine radiates on the order of 10$\%$ of its Eddington luminosity (Greenhill et al. 1997),
the estimated accretion rate from observations is 0.0031 M$_\sun$ yr$^{-1}$ which is less than the simulated mass 
inflow rate.  Our model can therefore basically provide sufficient mass inflow for the AGN of NGC 4945 at the 
50-pc scale.
However, we can see from the middle-right panel of Figure~\ref{vf} that the spiral waves become weaker inside 50 pc during the inward propagation.
The mass inflow rate would diminish accordingly.
Therefore, other mechanisms may be needed to drive the gas further inward to fuel the AGN on smaller distance scales.

\subsection{The importance of self-gravity of the gas}
Although it is known that self-gravity is unimportant in strong bar cases (Lindblad et al. 1996, Lin et al. 2008), we have carried out  
all the calculations described above with the inclusion of self-gravity 
because it may still play an important role for weak bars.
To understand its role in our calculations, we have also simulated the gaseous disk without self-gravity by  keeping all the settings identical to the case with self-gravity but turning off self-gravity.
The extent of the spirals becomes much smaller than that in the case with self-gravity.
This result is consistent with the asymptotic solutions in Yuan \& Cheng (1989) and Yuan \& Kuo (1997), who showed that self-gravity will shift the Q-barrier  toward the corotation region, and, as a result, the entire spiral pattern is shifted toward corotation, inward for the OLR and outward for the ILR.
Since the spiral pattern in our simulation lies inside the corotation, turning off self-gravity will move the spiral pattern inward and decrease its extent.
The morphology of the spirals are also different from that in the case with self-gravity.
Based on these tests, we believe that self-gravity should be included in the calculations for NGC 4945.

\subsection{Comparison between the hydrodynamical model and the gas orbit model}
In this subsection, we describe the differences between the hydrodynamic model presented in this paper and 
the gas orbit model presented in Lim et al. (2009).  In this latter work, an axisymmetric potential of the form 
$\Phi(x,y) = 1/2\ v_0^{2}\ ln(1+R^{2}/R_c^{2})$ was adopted. We note that such a potential is commonly adopted 
for only the galactic bulge part of the axisymmetric component of the gravitational potential. In addition, 
dissipation terms were introduced to characterize the damping of radial and azimuthal oscillations in the gas 
orbits which are assumed to be closed.  As noted by Lim et al. (2009), their model is therefore not self-consistent 
in at least these two aspects. Further specification of the model requires a description of their bar potential, 
whose radial profile is assumed to be the same as their axisymmetric component and with a perturbation amplitude 
corresponding to 5\% of that of the axisymmetric component.  Since the axisymmetric component they adopted is only 
the dominant part rather than the complete potential, the percentage of the pertubation to the complete axisymmetric 
potential should be less than 5\%, which indicates a weak bar.

In \S 2, we presented the functional form of the bar potential adopted in this paper (see eq. 7) as well as 
the motivation for choosing this particular form.  The strength of the bar potential $f$, defined in \S 2.2, 
is found to be 1.8\% which also corresponds to a weak bar.  Therefore, the two models are both based on a 
weak bar perturbation.

\begin{figure}
\figurenum{8}
\epsscale{0.8}
\plotone{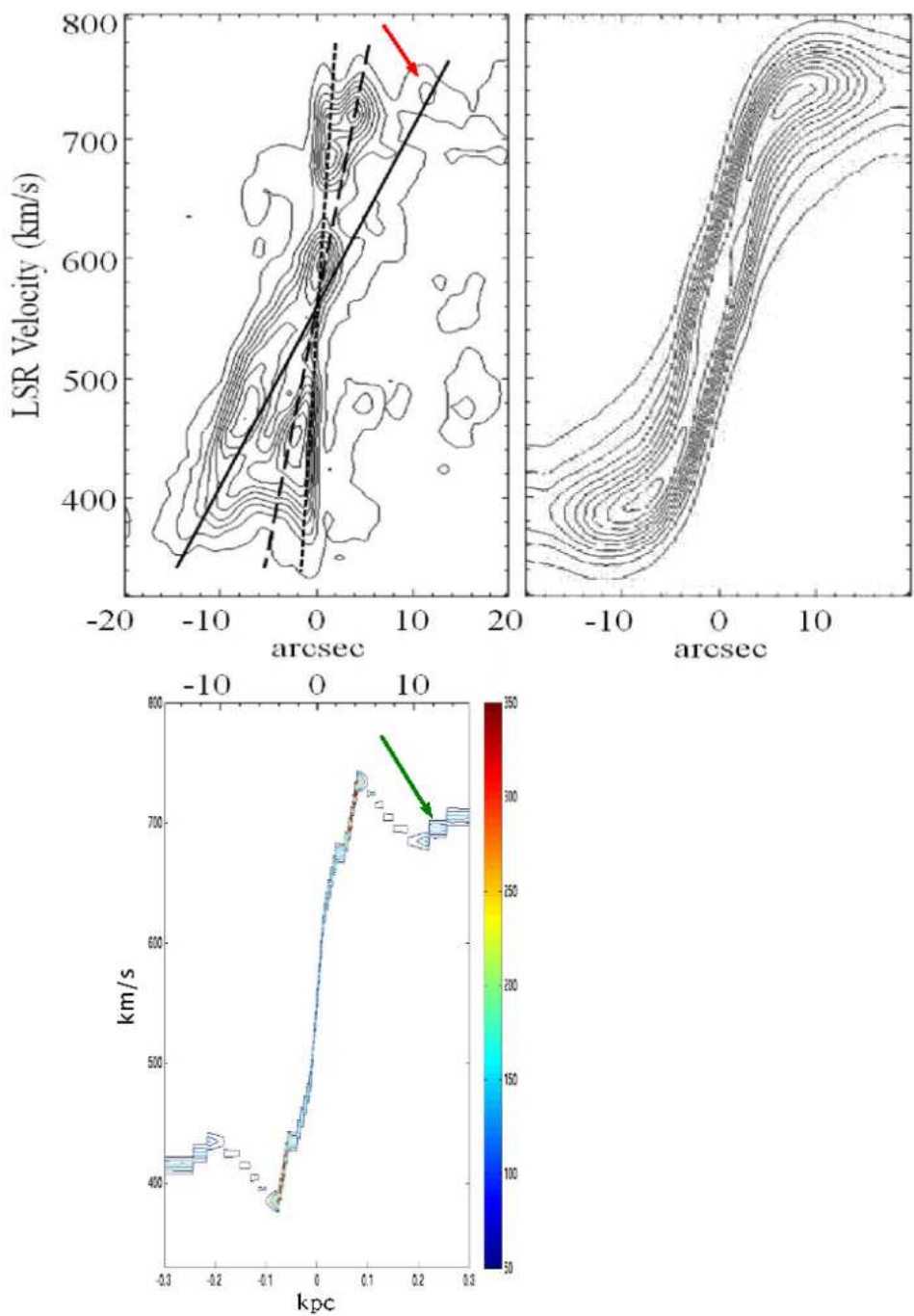}
\caption{
Upper-left: the observed P-V diagram in $^{12}$CO (J=2-1) in the left panel of Fig. 5 in Lim et al. (2009).
The intensity peak (marked by a red arrow) on the solid line is located at about 11.5$\arcsec$ ($\sim$ 220 pc).
Lower-left: the simulated P-V diagram. 
The color map denotes the surface density on a linear scale in units of M$_\sun$ pc$^{-2}$.
The density peak (marked by a green arrow) is at about 240 pc which almost coincides with that in the observation.
Upper-right: the position-velosity diagram of the gas orbit model in Fig. 15 in Lim et al. (2009). 
There is only one peak located at about 7$\arcsec$ ($\sim$ 140 pc).
}
\label{pv2}
\end{figure}

In the following, we compare the simulated P-V diagrams of these two models. For the observations, we 
make use of the P-V diagram in $^{12}$CO (J=2-1) shown in the left panel of Figure 5 in Lim et al. 
(2009), which is reproduced in the upper-left panel of Figure~\ref{pv2}.  This diagram, which is of a larger scale than the upper-right panel of Figure~\ref{pv},
reveals an additional feature not seen on the smaller scale of Figure~\ref{pv}. 
The short-dashed and long-dashed 
lines in this diagram have already been compared with those in our simulation in Figure~\ref{pv}.  The solid line 
indicates a third component which Lim et al. (2009) interpret as the central galactic molecular disk corresponding to the inner regions of the 
large-scale galactic disk.  
The intensity peak (marked by a red arrow) 
on the solid line is located at about 11.5$\arcsec$ ($\sim$ 220 pc).  The corresponding peak (marked by a green arrow) 
in our simulation (the lower panel of Figure~\ref{pv2}), produced by the outer parts of the spiral arms, is at about 240 pc which almost coincides with that in the 
observation.  On the other hand, Lim et al.(2009) claim that their model reproduces the gross kinematics of the 
central galactic molecular disk and molecular starburst ring, in the P-V diagram of the gas orbit 
model (Figure 15 in Lim et al. (2009) and reproduced in the upper-right panel of Figure~\ref{pv2} in this paper). 
However, even though they are able to obtain the outline of the kinematics, there is only one peak located at about 7$\arcsec$ ($\sim$ 140 pc), which does not match the position of 
any of the three peaks (1$\arcsec$, 4$\arcsec$, and 11.5$\arcsec$) in the observation.  Therefore, our model provides 
a better match to the observed P-V diagram than their gas orbit model.

The pattern speed of the bar potential in Lim et al. (2009) is 60 km s$^{-1}$ kpc$^{-1}$, according to which the 
inner and outer ILR's are located at $\sim$ 70 pc and  $\sim$ 1.2 kpc respectively.  In contrast, the pattern speed 
in our best-fit case is 233 km s$^{-1}$ kpc$^{-1}$ and the corresponding position of the outer ILR is at 0.19 kpc 
and there is no inner ILR.  Based on the comparisons presented in \S 3.2 with respect to Figure~\ref{pv} as well 
as above with respect to Figure~\ref{pv2}, we believe that the position of the outer ILR provided by our 
self-consistent model should be more reliable.

\section{Conclusions}
We have presented results of numerical simulations of the central region of NGC 4945 using the Antares hydrodynamics 
code which reproduce the main features of the velocity map of $^{12}$CO(2-1) and the map of the 1.3 mm continuum emission observed 
by Lim et al. (2009) using the SMA,
as well as the Pa$\alpha$ image of Marconi et al. (2000).  The S-shaped structure of the isovelocity contours in the 
SMA observation is well represented by the bending produced by a shock along the spiral density waves which are 
excited at the outer-inner Lindblad resonance by a rotating bar.  
Based on the comparisions between the results of our 
hydrodynamical simulations and the observations, we suggest that a small, weak, rapidly rotating bar could be present in 
the center of NGC 4945.
  
The simulated density map shows a pair of tightly wound spirals in the center passing through most of the ring-like 
high intensity region in the Pa$\alpha$ image.  This result provides an alternative interpretation to a starburst 
ring and does not necessarily imply the stalling of material at the ILR.  
That is, the presumed starburst rings need not be rings as they may be tightly wound spiral arms.

In terms of the overall structure, the observed velocity channel maps resemble the simulated maps.  
Comparisons between the observed and simulated P-V diagrams reveal that 
the observed compact and apparently rotating disk-like structures in NGC 4945 need not be disks and may,
instead, be unresolved spiral features. If these results are applicable to the general AGN population, then 
spiral features in the nuclear region of AGN may be more common than previously thought. The hydrodynamical 
model yields a mass inflow rate of 0.0085 M$_\sun$ 
yr$^{-1}$ at the 50 pc scale sufficient to fuel the AGN of NGC 4945.  We have also found that, although self-gravity 
is unimportant in strong bar cases, it is important for NGC 4945 and should be included in the model in order to 
obtain the best-fit simulation results.

We have compared our model with the gas orbit model discussed in Lim et al. (2009).  Both models require  
that the bar potential is weak, but our hydrodynamic model provides a better comparison to the observed P-V 
diagram.  Furthermore, while the gas orbit model predicts the presence of both the inner and outer ILR's, our 
result does not contain an inner ILR.  Given that the hydrodynamic model does not suffer the inconsistencies 
present in the gas orbit model adopted by Lim et al. (2009), the pattern speed and location of the outer ILR of the former model is to be 
preferred. 

Finally, we conclude by noting that in the complex central regions of galaxies it is common for observers to 
use P-V diagrams to select features of interest, often interpreting their distinguishing spatial-kinematic 
characteristics in terms of specific structures (e.g., disks and rings). However, if spirals 
are present even in the inner 10s of pc, as this paper suggests for NGC 4945, then one can not rely on simple 
gas orbit models. A proper treatment of the gas dynamics is necessary for detailed 
understanding. 

\acknowledgments
We thank the referee for his/her comments which helped to significantly improve the clarity and presentation of this work.
L.-H.L. and RET acknowledge the support of the Theoretical Institute for Advanced Research in Astrophysics 
(TIARA) operated under the Academia Sinica Institute of Astronomy \& Astrophysics in Taipei, Taiwan.

\clearpage

\begin{deluxetable}{llll}
\tabletypesize{\scriptsize}
\tablewidth{0pc}
\tablecaption{Adopted Bar Potential Parameters NGC 4945}
\tablehead{
\colhead{Parameter} &
\colhead{Self-gravitating disk} &
\colhead{Gas orbit model\tablenotemark{a}}
}
\startdata
Pattern speed $\Omega_p$ (km s$^{-1}$ kpc$^{-1}$) & 233 & 60 \\
Outer inner Lindblad resonance OILR (kpc) & 0.19 & 1.2 \\
Bar strength $f$ (\%)\tablenotemark{b} & 1.8 & 5\tablenotemark{c} \\
Location of the minimum of bar potential a$_1$ (kpc)\tablenotemark{d} & 0.09 & \\
\enddata
\tablenotetext{a}{This model is adopted in Lim et al. (2009).}
\tablenotetext{b}{$f$ is the ratio of the effective radial force exerted on the gaseous disk by the rotating bar (Yuan \& Kuo 1997) to the force for the circular motion from the rotation curve at the location of OILR.}
\tablenotetext{c}{The ratio of the amplitude of the bar potential to that of the axisymmetric component with the same radial profile (Lim et al. 2009).}
\tablenotetext{d}{a$_1$ denotes the length scale of the bar according to eq. 6 and 7.}
\end{deluxetable}
\clearpage

\end{document}